\documentclass[preprint,authoryear,review,12pt]{elsarticle}
\usepackage{amssymb}
\usepackage{lineno}
\setkeys{Gin}{draft=false}
\journal{Physics Earth Planetary Interiors}

\begin{document}

\linenumbers

\begin{frontmatter}

\title{Excitation of the Slichter mode by collision with a meteoroid or pressure variations at the surface
and core boundaries}

\author{S. Rosat}
\author{Y. Rogister}

\address{Institut de Physique du Globe de Strasbourg, UMR 7516 CNRS -- Universit\'e de Strasbourg (EOST) - 5 rue Descartes, 67084 Strasbourg, France}

\begin{abstract}
We use a normal-mode formalism to compute the response of a spherical, self-gravitating anelastic PREM-like Earth model 
to various excitation sources at the Slichter mode period. The formalism makes use of the theory of the Earth's free oscillations
based upon an eigenfunction expansion methodology. We determine the complete response in the form of Green's function obtained
from a generalization of Betti's reciprocity theorem. 
Surficial (surface load, fluid core pressure), 
internal (earthquakes, explosions) and external (object impact) sources of excitation are investigated to show that 
the translational motion of the inner-core would be best excited by a pressure acting at the core 
boundaries at time-scales shorter than the Slichter eigenperiods. 

\end{abstract}


\end{frontmatter}

\arraycolsep0.5pt

\section{Introduction}
The three free translational oscillations of the inner core, the so-called Slichter modes \citep{slichter}, have been a subject of observational controversy since the first detection by \citet{smylie} of a triplet of frequencies that he attributed to the Slichter modes. This detection has been supported by \citet{courtier} 
and \citet{pagiatakis} but has not been confirmed by other authors \citep{hinderer1995,jensen1995,rosat2006,guo2007}. 
Also, it motivated additional theoretical studies \citep{crossley,rochester,rieutord,rogister}. 
The search for the Slichter modes was invigorated by the 
development of worldwide data recorded by superconducting gravimeters (SGs) of the Global Geodynamics Project \citep{ggp}. 
Thanks to their long-time stability and low noise level, these relative gravimeters are the most suitable instruments 
to detect the small signals that would be expected from the Slichter modes \citep{hinderer1995, rosat2003, rosat2004}.

The theory is now better understood and computation predicts eigenperiods between 4 and 6 h \citep{rogister} for the seismological reference PREM \citep{prem}
 Earth model. A more recent study by \citet{grinfeldwisdom} states that the period could 
 be much shorter because of the kinetics of phase transformations at the inner-core boundary (ICB).

The observation of the Slichter modes is fundamental because, the restoring force being Archimedean, their periods are directly related to the density jump at the ICB. 
This parameter is still poorly known: 
by analyzing seismic PKiKP/PcP phases, \citet{koperpyle} found that it should be smaller than 450 kg/m$^3$, later increased to 520 kg/m$^3$ \citep{koperdombro}, 
whereas \citet{mastersgubbins} obtained 820 $\pm$ 180 kg/m$^3$ from normal modes observation. \citet{tkalcic} have shown that the uncertainties associated 
with the seismic noise might partially explain such discrepancies for the estimates of the ICB density contrast. \citet{gubbins} have proposed a model with a large overall density jump between 
the inner and outer cores of 800 kg/m$^3$ and a sharp density jump of 600 kg/m$^3$ at the ICB itself. Such a model satisfies both the constraints set by powering the geodynamo 
with a reasonable heat flux from the core, and PKP traveltimes and normal mode frequencies. The value of the density jump at ICB for the PREM model is 600 kg/m$^3$.

This paper aims at evaluating the possible amplitude of the Slichter modes for various types of excitation sources. 

The seismic excitation has been previously studied by \citet{smith}, \citet{crossley} and \citet{rosat2007}. They have shown that the best natural focal 
mechanism to excite the Slichter mode is a vertical dip-slip source. The largest magnitude event in the past was 
the 1960 Chile earthquake with a magnitude $M_w=9.6$ for the main shock. A foreshock occurred with a magnitude of 9.5 \citep{kanamori}. 
The combination of both events leads to a seismic source of magnitude $M_w=9.8$ which would be enough to excite the Slichter modes to the nanoGal level. 
However, at such frequencies, the noise levels of SGs are of several nGal, even for the quietest sites \citep{rosathinderer}.
Earthquakes are therefore not the most suitable source to excite the Slichter modes to a level sufficient for the SGs to detect the induced surface gravity effect. 

Surficial pressure flow acting in the core has been considered by \citet{greff} as a possible excitation source.
In this work, we reconsider the pressure flow acting in the core using a Green function formalism for a non-rotating 
anelastic PREM Earth model. Then, we investigate the surface load and meteoroid impact as possible sources of excitation of the Slichter modes.

\section{Green function formalism}
\label{GreenFunctionFormalism} 
We consider a spherical non-rotating anelastic Earth model. 
The displacement $\mathbf u$ at a point $\mathbf r$ and time $t$ produced by any body force density ${\mathbf f}$ acting in volume $V$ and surface 
force density ${\mathbf t}$ acting upon surface $S$ can be written as a convolution of the impulse response $G$ with the entire past 
history of the forces ${\mathbf f}$ and ${\mathbf t}$ \citep{DT}:
\begin{eqnarray}
{\mathbf u} ({\mathbf r},t)=\int_{-\infty}^{t}\int_{V} G({\mathbf r},{\mathbf r}';t-t'){\mathbf f}({\mathbf r}',t')dV'dt' + 
 \int_{-\infty}^{t}\int_{S} G({\mathbf r},{\mathbf r}';t-t'){\mathbf t}({\mathbf r}',t')d\Sigma'dt', \nonumber \\
\label {uG}
\end{eqnarray}
where ${\mathbf r}'$ is the integrated position vector. 
This relation is inferred from Betti's reciprocity relation in seismology \citep{aki}.
Seismic Green's tensor $G$ of a non-rotating anelastic Earth is given in terms of the normal-mode complex frequencies $\nu_k = \omega_k(1+\frac{i}{2Q_k})$ and eigenfunctions ${\mathbf s}_k$ by
\begin{equation}
G({\mathbf r},{\mathbf r}';t)=\Re \sum_{k} (i \nu_k)^{-1}{\mathbf s}_k({\mathbf r}){\mathbf s}_k({\mathbf r}')e^{i\nu_k t}H(t),
\label{GreenFunction}
\end{equation}
where $\Re$ denotes the real part of the complex expression and $H(t)$ is the Heaviside function.

\citet{tromp} have generalized Betti's reciprocity relation to a representation theorem suited for surface-load problems, 
so that the displacement ${\mathbf u}$ due to a surface load $\sigma$ located at ${\mathbf r}'$ is given by
\begin{equation}\label{uGg}
{\mathbf u}({\mathbf r},t)=\int_{-\infty}^{t}\int_{S} \sigma({\mathbf r}',t') {\mathbf \Gamma} ({\mathbf r},{\mathbf r}';t-t')d\Sigma'dt',
\end{equation}
where $\mathbf \Gamma$ is the surface-load Green's vector defined by
\begin{equation}\label{gamma}
{\mathbf \Gamma}({\mathbf r},{\mathbf r}';t)=-[G({\mathbf r},{\mathbf r}';t) \cdot \nabla'\Phi({\mathbf r}')+{\mathbf g}({\mathbf r},{\mathbf r}';t)], 
\end{equation}
$\nabla'$ is the gradient with respect to ${\mathbf r}'$, $\Phi$ is the unperturbed gravitational potential, and ${\mathbf g}$ is
\begin{equation}
{\mathbf g}({\mathbf r},{\mathbf r}';t)=\Re \sum_{k} (i \nu_k)^{-1}\phi_k({\mathbf r}){\mathbf s}_k({\mathbf r}')e^{i\nu_kt}H(t), 
\label{gVector}
\end{equation}
$\phi_k$ denoting the perturbation of the gravitational potential 
associated with the normal mode $\{ {\mathbf s}_k,\phi_k \}$.
Green's tensor $\{G,{\mathbf g}\}$ represents the complete point-source response.

A spheroidal mode of harmonic degree $l$ and order $m$ and radial overtone number $n$ 
can be decomposed into three components in spherical coordinates:
\begin{equation}\label{slm}
_n {\mathbf s}_l^m({\mathbf r})={ _nU_l^m}(r)Y_l^m(\theta,\phi)\hat {\mathbf r}+k^{-1} {_nV_l^m}(r)\frac{\partial Y_l^m}{\partial \theta}\hat {\mathbf \theta}+
k^{-1} {_nV_l^m}(r)\frac{1}{\sin{\theta}}\frac{\partial Y_l^m}{\partial \phi}\hat {\mathbf \phi},
\label{displacement}
\end{equation}
where $Y_l^m (\theta,\phi)$ are the real spherical harmonics of degree $l$ and order $m$ \citep{DT}, 
$k=\sqrt{l(l+1)}$ and $\hat {\mathbf r}$, $\hat {\mathbf \theta}$ and $\hat {\mathbf \phi}$ 
are the usual unit vectors of the spherical coordinates. The associated perturbation of the gravitational potential has the form 
\begin{equation}
_n\phi_l^m({\mathbf r})= {_nP_l^m}(r)Y_l^m(\theta,\phi).
\end{equation}
The eigenfunctions $_nU_l^m(r)$, $_nV_l^m(r)$ and $_nP_l^m(r)$ are functions of the radius only. 
Because the model is non-rotating and spherically symmetric, the $2l + 1$ eigenfrequencies 
for each fixed $l$ and $n$ are degenerate into a single eigenfrequency that we can therefore denote by $_n\nu_l$.
The summation over $k$ in Eqs (\ref{GreenFunction}) and (\ref{gVector}) is actually a triple summation over $l$, $m$ and $n$.
Since the eigenfrequencies do not depend on $m$, 
the summation over $m$ can be performed using the addition theorem for surface spherical harmonics, with the result [\citet{DT}, Eqs 10.28 and 10.34]:
\begin{eqnarray}
G({\mathbf r},{\mathbf r}';t) 
= \Re \sum_{n}\sum_{l} & &  \frac{2l+1}{4\pi} \frac{e^{i _n\nu_lt} }{i _n\nu_l}\nonumber \\ 
& & \{ _nU_l(r) _nU_l(r')\hat {\mathbf r}\hat {\mathbf r}'P_{l0} \nonumber \\
& & + k^{-1}[ _nU_l(r) _nV_l(r')\hat {\mathbf r}\hat {\mathbf \Theta'}- _nV_l(r) _nU_l(r')\hat {\mathbf \Theta}\hat {\mathbf r}']P_{l1} \nonumber \\
& & + \frac{1}{2}k^{-2}[ _nV_l(r) _nV_l(r')\hat {\mathbf \Theta}\hat {\mathbf \Theta'}](k^2 P_{l0}-P_{l2}) \nonumber \\
& &+ k^{-2}[ _nV_l(r) _nV_l(r')\hat {\mathbf \Phi}\hat {\mathbf \Phi'}](\sin\Theta)^{-1}P_{l1}\},
\label{GreenSphericalModel}
\end{eqnarray}
where $\hat {\mathbf \Phi}=\hat{\mathbf r}\times \hat{\mathbf \Theta}$ and 
$\Theta$ is the angular distance between the receiver at $\hat{\mathbf r}$ and the point source at $\hat{\mathbf r}'$ :
\begin{equation} 
\cos \Theta=\hat{\mathbf r} \cdot \hat{\mathbf r}'=\cos \theta \cos \theta'+\sin \theta \sin \theta'\cos(\phi-\phi').
\label{bigtheta}
\end{equation}

\begin{figure}[!ht]
\centering
\noindent\includegraphics[width=15cm]{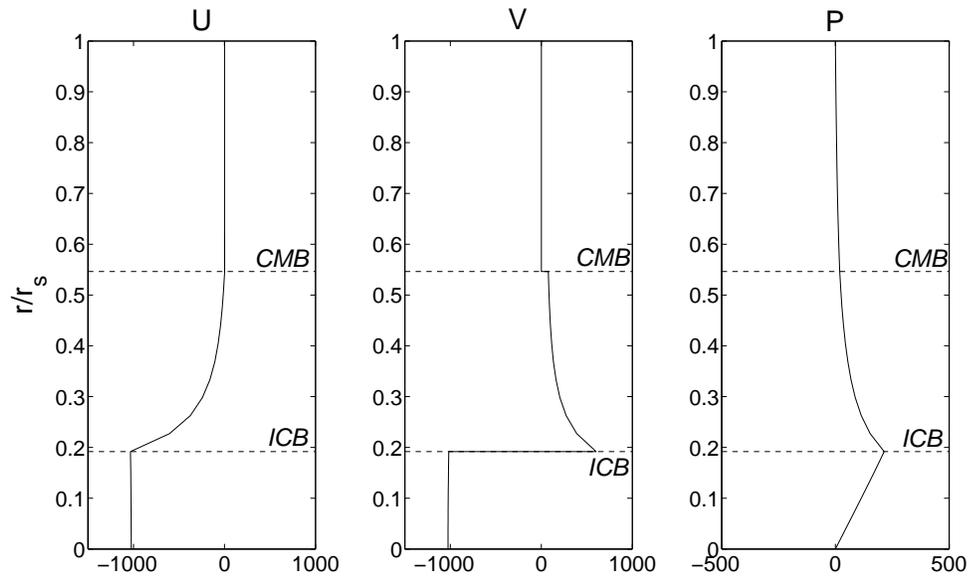}
\caption{Eigenfunctions of the Slichter mode $_1{\rm S}_1$ for the PREM model. 
The vertical axis is the radius normalized by the Earth's surface radius $r_s$. 
$U$ and $V$ are the radial dependence of the displacement given 
by Eq. (\ref{displacement}). 
$P$ is the perturbation of the gravitational potential. 
The normalization of the eigenfunctions is such that $U(r_s) = 1$ m.
}
 \label{fig:UVP}
 \end{figure}

The Slichter mode is the spheroidal mode of harmonic degree one and radial overtone number one. 
For a non-rotating spherical model, the three Slichter frequencies are degenerate into a single 
eigenfrequency.
As 
\begin{equation}
P_{10}(\cos\Theta)=\cos\Theta
\end{equation} 
and 
\begin{equation}
P_{11}(\cos\Theta)=\sin\Theta, 
\end{equation}
the term for which $l=1$ and $n=1$ in Eq. (\ref{GreenSphericalModel}) writes:
\begin{eqnarray} \label{Gl}
_1G_1 ({\mathbf r},{\mathbf r}';t) = \frac{3}{4\pi} \Re & & \lbrace
\frac{e^{i \nu} }{i \nu}
[U(r)U(r')\hat{\mathbf r}\hat{\mathbf r}'\cos\Theta \nonumber \\ 
& &+ \frac{1}{\sqrt{2}}(U(r)V(r')\hat{\mathbf r}\hat{\mathbf \Theta'}-V(r)U(r'){\mathbf \Theta}\hat{\mathbf r}')\sin\Theta \nonumber \\ 
& & +\frac{1}{2}V(r)V(r')\hat{\mathbf \Theta}{\mathbf \Theta'}\cos\Theta+\frac{1}{2}V(r)V(r')\hat{\mathbf \Phi}\hat{\mathbf \Phi'}]\rbrace
\label{Green_one_one}
\end{eqnarray}
being understood that $\nu = {_1\nu_1}$, $U = {_1U_1}$, $V = {_1V_1}$ and $P = {_1P_1}$. 
For PREM, the Slichter eigenperiod is 5.42 h \citep{rogister} and the eigenfunctions $U$, $V$ and $P$ are plotted in Fig. \ref{fig:UVP}. 

The damping rate depends on the dissipation processes involved. 
A summary of plausible dissipation processes is given by \citet{greff, guo2007, rosatCDW}. 
The role of the outer core viscosity has been studied by \citet{smyliemc} and \citet{rieutord}, the effect of a mushy zone at the ICB, by \citet{peng}, the influence
of the magnetic field, by \citet{buffett} and the anelastic dissipation for the core modes, by \citet{crossleyetal}. 
Such studies have revealed that it is unlikely that the damping factor of the Slichter mode be less than 2000, corresponding to a damping time of 144 days. 
In this case, the induced surface gravity perturbation should be more easily detectable by SGs. 
We assume, in the following, a quality factor of 2000. 

Using the Green function formalism, we can compute the excitation of the Slichter mode by any body or surface forces.

\section{Excitation by fluid core pressure}\label{fluidcore}

Observational evidence for motions in the core comes from core-sensitive seismic modes, which have periods smaller than one hour, 
the free core nutation, which is a rotational mode of nearly-diurnal period, and variations of the magnetic field that can be related to motions in the core with 
timescales larger than one year. 
Therefore, the dynamics of the fluid core at the Slichter frequencies lacks observational evidence. 
Theoretical results suggest that, at timescales smaller than one day and outside the seismic band, plausible motions 
are to be searched for in the turbulent convection or in the spectrum of the core. 
 
An account of small-scale turbulence driven by convection is given by \citet{loper}. 
The timescale may be less than one day but, because of the small characteristic length-scales, 
turbulence is unlikely to excite the translation of the whole inner core.
 
\citet{valettea,valetteb} has shown that the inertia-gravity spectrum of an inviscid liquid core 
is continuous and set bounds on it. The bounds depend on both the speed of rotation and squared Brunt-V\"ais\"al\"a 
frequency. 
\citet{rogvalette} and \citet{rogister2010} have suggested that the 
rotational modes might be influenced by the continuous spectrum in which they are embedded. 
In particular, the nearly-diurnal free inner core nutation and long-period inner core wobble 
might be double or even multiple and have energy in the liquid core. 
Pending on the value of the squared Brunt-V\"ais\"al\"a frequency in the outer core, 
the Slichter modes could also be embedded in the continuous spectrum. 
Similarly to what has been found for the two rotational modes of the inner core, 
significant motion and pressure variation in the liquid core could accompany 
the Slichter modes. 
Although a Slichter mode with its associated motion in the liquid core should then be 
considered as a single normal mode, 
we can for simplicity assume that the pressure variations in the liquid core excite the 
translational motions of the inner core. 
This is somewhat the opposite of what \citet{buffett2010} did to investigate the attenuation 
of the free inner core nutation: 
he assumed that the tilt of the inner core generates shear layers in the outer core 
where Ohmic and viscous dissipation occur.

As \citet{greff}, we will assume that the pressure at the CMB takes the following analytical form:
\begin{equation}
P^c(\theta, \, \phi, \, t) = P_0^c(\theta, \, \phi) e^{-(\frac{t-T_0}{\tau})^2},
\end{equation}
where  $T_0$ is the starting time of application of the pressure, $\tau$ is 
the time duration of the pressure source and
$P_0^c(\theta, \, \phi)$ includes three terms of harmonic degree 1:
\begin{equation}
P_0^c(\theta, \, \phi) =P_{10}^c \cos\theta + (P_{11}^c \cos\phi + {\tilde P}_{11}^c \sin\phi) \sin\theta .
\end{equation}
According to \citet{okubo} and \citet{greff}, the total force exerted at the core boundaries must vanish 
for the centre of mass to be kept fixed. 
This translates into the so-called Consistency Relation \citep{farrell} and imposes a relation between 
the pressures $P^c$ and $P^{ic}$ at the CMB and ICB, respectively:
\begin{equation}
P^{ic} = \frac{r_c^2}{r_{ic}^2} P^c.
\label{Pic}
\end{equation} 

\citet{greff} analytically solved the equations of the elasto-gravitational deformation \citep{alterman} 
for a simple Earth model made up of three homogeneous incompressible layers and 
investigated the excitation of the Slichter mode by a pressure acting at the outer core boundaries. 
In this section, we consider the same simple excitation sources to test the Green function approach 
for PREM, which is a more realistic Earth model. 
We compute the displacement ${\mathbf u}$ by means of Eqs (\ref{uG}) and (\ref{Green_one_one}). 
As we are mainly interested in the surface gravity effect, 
we only need the radial component of the displacement: 
\begin{eqnarray}
& & u_r(r,\theta,\phi,t) = \frac{3}{4\pi} \Re \lbrace \frac{U(r)}{i\nu} 
 \int_{-\infty}^{t} e^{i\nu (t-t')} e^{-[\frac{t'-T_0}{\tau}]^2} dt' \nonumber \\
& & \hspace{5mm}  \lbrack U(r_c) \int_{\rm CMB} \cos\Theta (P_{10}^c \cos\theta' 
+ (P_{11}^c \cos\phi' + {\tilde P}_{11}^c \sin\phi') \sin\theta') d\Sigma' \nonumber \\
& & \hspace{5mm}  - U(r_{ic}) \int_{\rm ICB} \cos\Theta (P_{10}^{ic} \cos\theta'  
+ (P_{11}^{ic} \cos\phi' + {\tilde P}_{11}^{ic} \sin\phi') \sin \theta') d\Sigma' \rbrack \rbrace 
\end{eqnarray}
At the CMB, $d\Sigma' = r_c^2 \sin\theta' d\theta' d\phi'$ and, at the ICB, $d\Sigma' = r_{ic}^2 \sin\theta' d\theta' d\phi'$. 
Taking Eqs (\ref{bigtheta}) and (\ref{Pic}) into account, the integration over $\theta'$ and $\phi'$ gives:
\begin{eqnarray}
 u_r(r,\theta,\phi,t)
 =  \Re \lbrace && \frac{r_c^2 U(r)}{i\nu} \lbrack U(r_c)-U(r_{ic})\rbrack 
 \int_{-\infty}^{t} e^{i\nu (t-t')} e^{-[\frac{t'-T_0}{\tau}]^2} dt' \nonumber \\
&&(P_{10}^c \cos\theta + P_{11}^c \sin\theta \cos\phi + {\tilde P}_{11}^c \sin\theta \sin\phi) \rbrace
\end{eqnarray}
The integral 
\begin{equation}
I(t) =  \int_{-\infty}^{t} e^{i\nu (t-t')} e^{-[\frac{t'-T_0}{\tau}]^2} dt'
\label{Time_Integral}
\end{equation}
is calculated in the Appendix. 
It gives 
\begin{equation}
I(t)=\frac{\sqrt{\pi}}{2}\tau e^{i\nu(t-T_0)}e^{-\nu^2\tau^2/4}[1+ \mbox{erf}(\frac{t-T_0}{\tau}+i\frac{\nu \tau}{2})],
\end{equation}
where $\mbox{erf}$ denotes the error function.
The radial displacement then becomes:
\begin{eqnarray}
u_r(r,\theta,\phi,t)& =& 
-\frac{r_c^2 U(r)}{\omega(1+\frac{1}{4Q^2})}[U(r_c)-U(r_{ic})] 
 \lbrack \frac{\Re\{I(t)\}}{2Q}-\Im\{I(t)\} \rbrack  \nonumber \\
& &  (P_{10}^c \cos\theta + P_{11}^c \sin\theta \cos\phi + {\tilde P}_{11}^c \sin\theta \sin\phi),
\end{eqnarray}
where $\Re \{ \, \}$ and $\Im \{ \, \} $ respectively denote the real and imaginary parts 
of the expression between brackets.

The degree-$l$ gravity variation measured by a gravimeter at the surface of the Earth $r_s$ is the sum of three terms: 
the free-air gravity variation owing to the displacement of the ground in the surrounding 
unperturbed gravity field $g_0$
\begin{equation}
g_{\rm{free}}=(-4\pi G\bar{\rho}+\frac{2}{r_s}g_0)U(r_s),
\end{equation}
the inertial acceleration of the ground 
\begin{equation} 
g_{\rm in}=-\omega_l^2U(r_s) 
\end{equation}
and the perturbation of the gravitational attraction 
\begin{equation}
 g_{\rm pot}=4\pi G \bar{\rho} U(r_s)+\frac{2}{r_s}P(r_s).
\end{equation}
In these expressions, $\bar{\rho}$ is the mean density of the Earth.

The degree-1 gravity variation measured by a gravimeter is therefore
\begin{eqnarray}
\Delta g(\theta, \, \phi, \, t) & = & \frac{r_c^2}{\omega(1+\frac{1}{4Q^2})}[U(r_c)-U(r_{ic})]
\nonumber \\ 
& &[P_{10}^c \cos\theta 
+ P_{11}^c \sin\theta \cos\phi + {\tilde P}_{11}^c \sin\theta \sin\phi]  \nonumber \\
& & \lbrack \frac{\Re\{I(t)\}}{2Q} -\Im\{I(t)\} \rbrack 
\lbrack -\omega^2U(r_s)+\frac{2}{r_s}g_0 U(r_s)+\frac{2}{r_s}P(r_s) \rbrack
\end{eqnarray}

\begin{figure}[!ht]
\centering
\noindent\includegraphics[width=15cm]{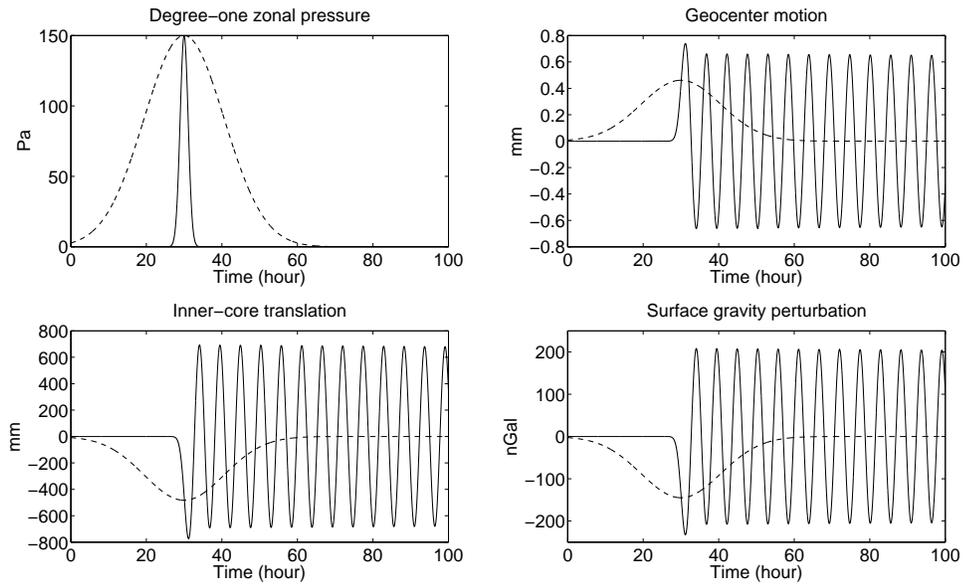}
\caption{Effects of the excitation of the Slichter mode 
by a fluid pressure acting at the CMB for two different excitation time-scales: $\tau=$ 1.5 h (solid line) and $\tau=$ 15 h (dashed line).
This figure is similar to Fig. 3 of \citet{greff} but we have applied a Green tensor formalism
to the PREM model. (a) Degree-one zonal pressure; (b) geocentre motion; (c) inner-core translation;
 (d) surface gravity perturbation.}
 \label{fig:pressICB}
 \end{figure}

We consider a zonal pressure $P_{10}^c =$ 150 Pa and compute the induced geocentre motion, inner-core translation and surface gravity perturbation for both
$\tau =$ 1.5 h and 15 h (Fig. \ref{fig:pressICB}). 
As the centre of mass is fixed, the geocentre motion, which is the displacement of the figure centre 
with respect to the centre of mass, corresponds to the surface displacement \citep{greff1997}.

\begin{figure}[!ht]
 \centering
\noindent\includegraphics[width=15cm]{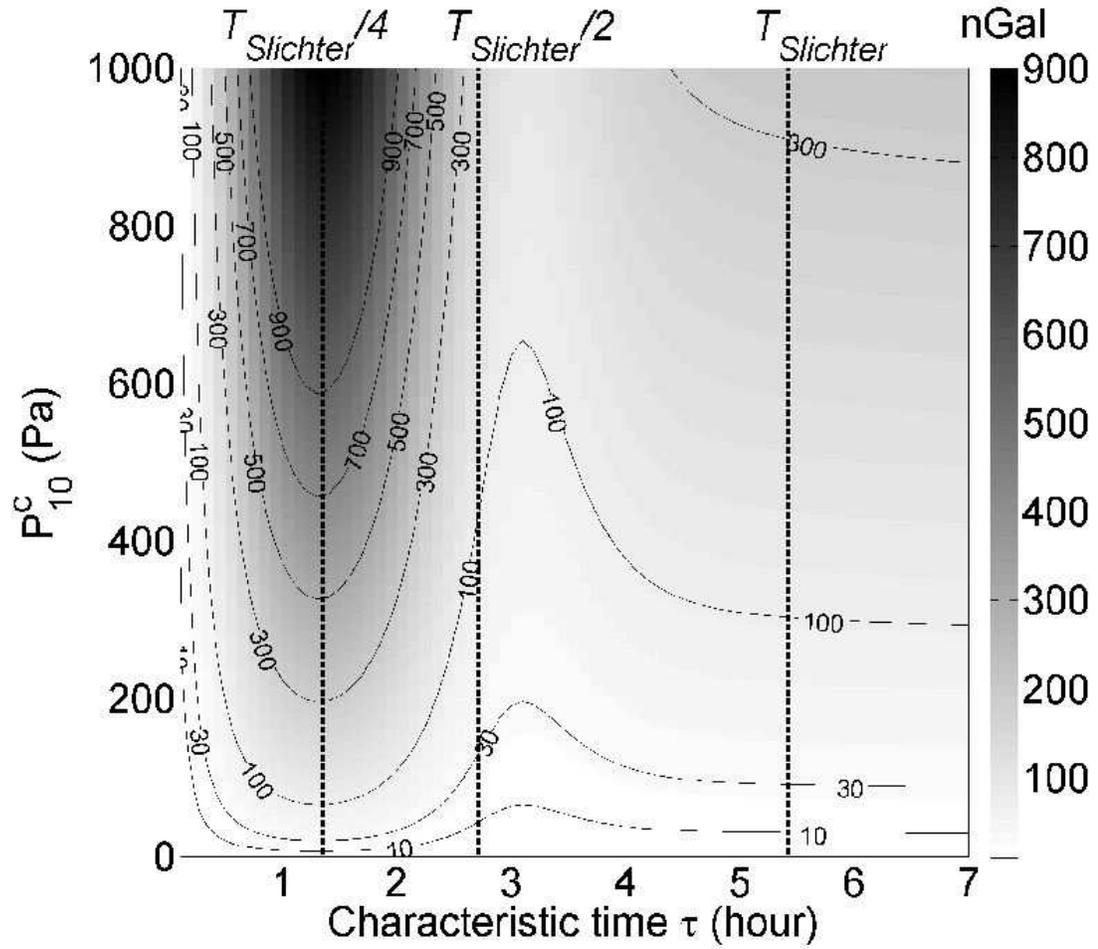}
 \caption{Surface gravity perturbation induced by the Slichter mode excited 
 by a fluid pressure acting at the CMB for different excitation time-scales and various zonal pressure amplitudes. 
 The vertical dotted lines correspond to one fourth of the Slichter period, 
 one half of the Slichter period and the Slichter period.}
 \label{fig:pressICB2D}
 \end{figure}

For  $P_{10}^c$ varying from 0 to 1000 Pa and $\tau$ ranging from 0.1 and 10 h, we compute a 2D-map (Fig. \ref{fig:pressICB2D}) of the surface gravity perturbation 
when the exciting pressure has vanished (i.e. $t>T_0+\tau$).

Our results for the PREM model show excitation amplitudes larger by 
70 $\%$ than the amplitudes computed by \citet{greff}, who used a very 
simple model made up of three incompressible homogeneous layers with a 
solid inner core, a liquid outer core and a rigid mantle. 
If we consider such a 3-layer model with the average densities of the 
PREM inner core, outer core and mantle, 
we obtain a period of 3.09 h for the Slichter mode and excitation amplitudes 
in close agreement with the values of \citet{greff}. 
They do not provide numerical details for 
the structure of their model but mention that its Slichter period is 3.08 h, 
which is almost equal to the Slichter period of our 3-layer model. 
Therefore, the Green function formalism we have adopted 
gives the same result as the analytical solution for the degree-1 deformation of 
a simple 3-layer model obtained by \citet{greff}.

The difference between the excitation amplitudes we obtain for PREM 
and the amplitudes computed by \citet{greff} comes from the different 
Earth models. 
The elasticity of the mantle and inner core, the compressibility of the outer core 
and the density jump at the ICB all come into play.
\citet{greff} mention that the elasticity of the mantle perturbs the solutions by 30 $\%$, 
without any further specifications.
We have checked that, by making the outer core of our simple 3-layer model 
compressible, with a P-wave velocity of 11083 m/s, the excitation amplitude of 
the Slichter mode is decreased by 40 $\%$ with respect to the incompressible
model.
The influence of the compressiblity and stratification of the core 
on the Slichter mode was also investigated by \citet{rogister}.

The perturbation of the surface gravity field is the largest when $\tau$ is 
smaller than half the Slichter eigenperiod. 
When $\tau$ is one fourth of the Slichter period, a 10 Pa pressure acting at the CMB, 
which by Eq. (\ref{Pic}) imposes a 81 Pa pressure at the ICB, is enough to induce a 
10 nGal (0.1 nm/s$^2$) surface gravity perturbation, which should be detectable by SGs.

\section{Excitation by a surface load}
A degree$-l$ surficial mass distribution $\sigma^s$ at the surface $r_s$ exerts forces 
over the Earth in two ways. 
First, at the interface between the Earth and the load $\sigma^s$, the static contact forces give rise to a degree$-l$ pressure 
\begin{equation}
P^s = g_0\sigma^s.
\end{equation}
Second, the gravitational attraction of the load $\sigma ^s$ over the entire Earth 
is described by a degree$-l$ potential
\begin{equation}
\phi = \frac{4\pi G}{2l+1}\sigma^s r_s \left \{ \begin{array}{ll} 
(\frac{r}{r_s})^l & \textrm{if r $\leq$ $r_s$} \\
(\frac{r_s}{r})^l & \textrm{if r $>$ $r_s$}
\end{array} \right.
\end{equation}

Atmospheric pressure models are sampled at 3 h at best. So instead of using actual data, 
we write the surface density load in the analytical form:
\begin{equation}
{\sigma}^s (\theta, \, \phi, \, t)=\sigma_0^s(\theta, \, \phi) e^{-[\frac{t-T_0}{\tau}]^2},
\end{equation}
which is the same as the expression used for the fluid core pressure in Section \ref{fluidcore}. 
The degree-one load $\sigma_0^s$, like $ P_0^c$,  contains three terms:
\begin{equation}
\sigma_0^s(\theta, \, \phi) =\sigma_{10} \cos\theta + (\sigma_{11} \cos\phi + {\tilde \sigma}_{11} \sin\phi) \sin\theta .
\end{equation}

After integration of Eq. (\ref{uGg}) over the whole surface, the radial displacement is given by:
\begin{eqnarray}
u_r(r,\theta,\phi,t) & = & \frac{r_s^2 U(r)}{\omega(1+\frac{1}{4Q^2})} [U(r_s)g_0+P(r_s)] 
\lbrack \frac{\Re\{I(t)\}}{2Q}-\Im\{I(t)\} \rbrack \nonumber \\ 
& & \lbrack  \sigma_{10} \cos\theta + \sigma_{11} \sin\theta \cos\phi 
 + {\tilde \sigma}_{11} \sin\theta \sin\phi \rbrack ,
\end{eqnarray}
and the perturbation of the surface gravity is
\begin{eqnarray}
\Delta g(t)&=& \frac{r_s^2}{\omega(1+\frac{1}{4Q^2})}[U(r_s)g_0+P(r_s)] 
\lbrack \frac{\Re\{I(t)\}}{2Q} -\Im\{I(t)\} \rbrack \nonumber \\ 
& & [\sigma_{10} \cos\theta  + \sigma_{11} \sin\theta \cos\phi + {\tilde \sigma}_{11} \sin\theta \sin\phi]  \nonumber \\  
& & [-\omega^2U(r_s)+\frac{2}{r_s}g_0 U(r_s)+\frac{2}{r_s}P(r_s)]
\end{eqnarray}

\begin{figure}[!ht]
\centering
\noindent\includegraphics[width=15cm]{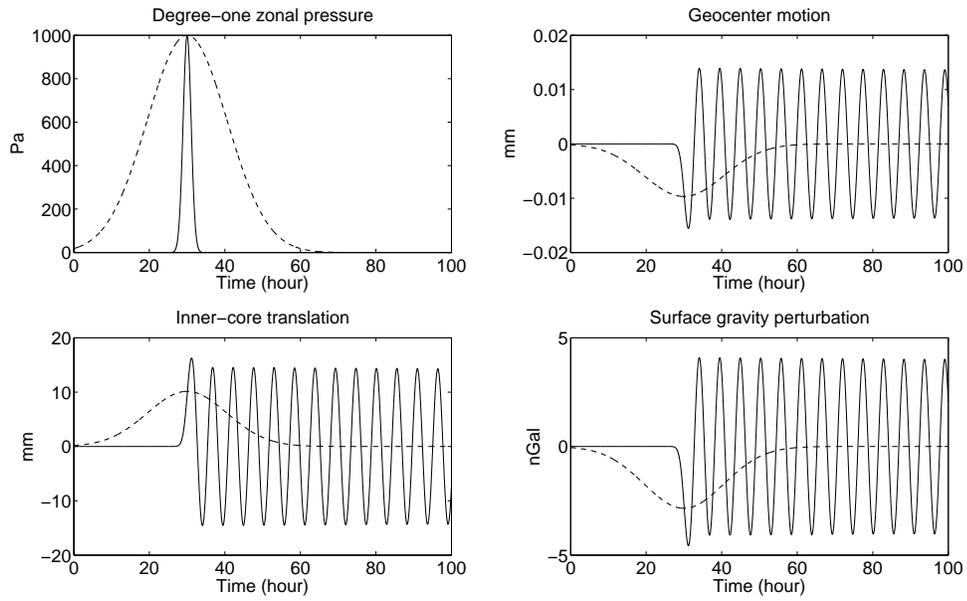}
\caption{Resulting effects of the excitation of the Slichter mode 
by a zonal surface load $\sigma(t)=\sigma_0 e^{-[\frac{t-T_0}{\tau}]^2}$ for two 
different excitation time-scales: $\tau=$ 1.5 h (solid line) and $\tau=$ 15 h (dashed line).
(a) Degree-one zonal pressure effect; (b) geocentre motion; (c) inner-core translation;
(d) surface gravity perturbation.}
\label{fig:surfaceload}
\end{figure}

We use a zonal surface load pressure of 1000 Pa (the surface mass density is then $\sigma_{10}=P_{10}/g_0$) 
and we compute the induced geocentre motion, the inner-core translation and the surface gravity perturbation for 
two excitation time-scales (1.5 h and 15 h) (Fig. \ref{fig:surfaceload}). 
When applying a surface load of 1000 Pa during $2\tau = 3$ h, the induced surface gravity perturbation 
has an amplitude of 5 nGal (0.05 nm/s$^2$) corresponding to an inner-core translation of 15 mm and a 
geocentre motion in the opposite direction with an amplitude of 0.015 mm. 
When the excitation time-scale (15 h) is larger than the Slichter period, the excitation amplitude is smaller.

\begin{figure}[!ht]
\centering
\noindent\includegraphics[width=15cm]{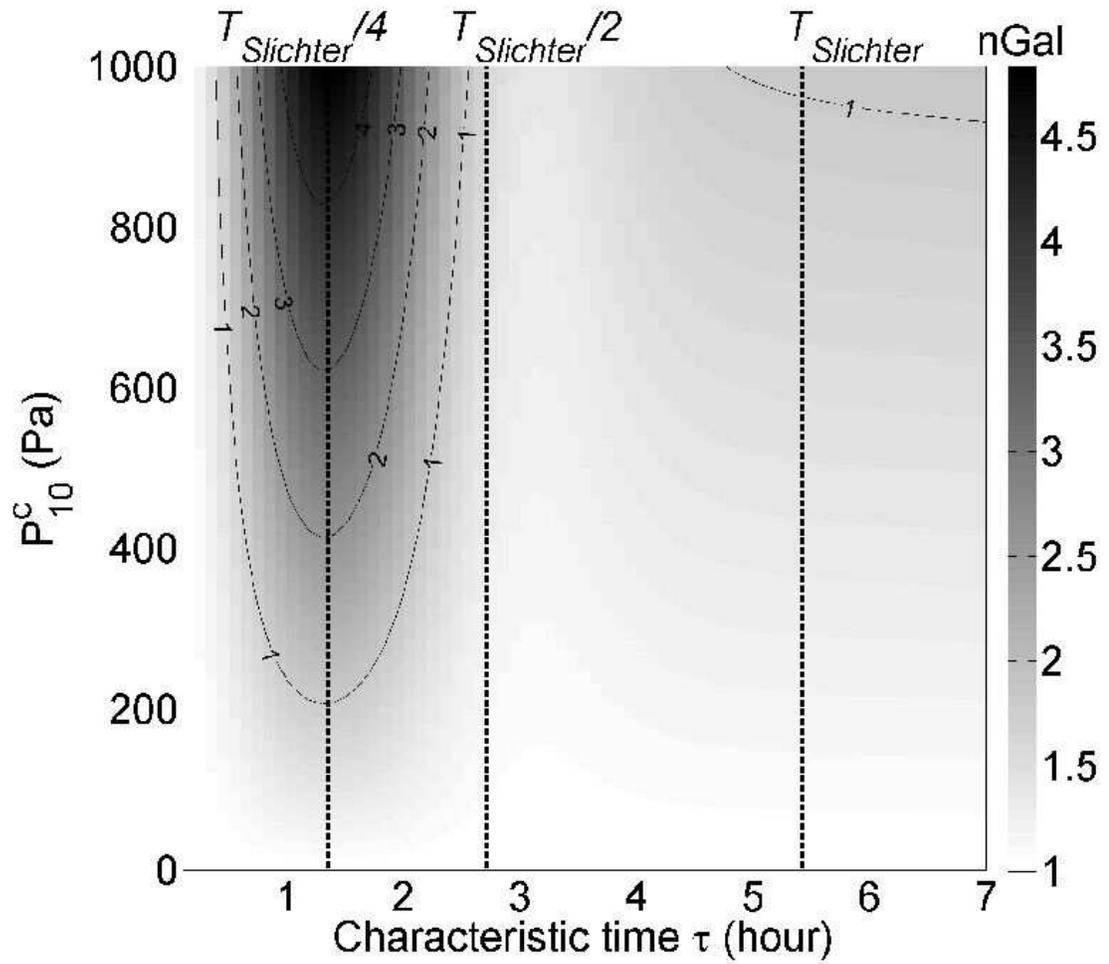}
\caption{Surface gravity perturbation induced by the Slichter mode excited 
by a surface load for different excitation time-scales and various zonal pressure amplitudes.}
\label{fig:surfaceloadgravi2D}
\end{figure}

We also plot the surface gravity perturbation associated to the Slichter mode excited 
by a surface load for different excitation time-scales and various zonal pressure amplitudes in Fig. \ref{fig:surfaceloadgravi2D}.
The conclusion is similar to the one for an internal pressure flow, except that the surface gravity 
variations are about 300 times smaller.

In this section, we have estimated the effect of a surface load on the Slichter mode. 
Another source that, intuitively, could make the inner core oscillate is a shock at the surface. 
Hence, in the next section, we study the excitation by an object impact on the Earth's surface.

\section{Excitation by an object impact}
Stellar objects, such as asteroids or comets, are dragged by the Earth's atmosphere and reach the Earth's surface at relatively modest velocities, typically a few tens of km/s. 
The released energy is comparable to nuclear explosions (according to Table 6 of \citet{collins}, from 3.2 to 3.9 $10^8$ megatons of TNT, where 1 Mt $=$ 4.2 $10^{15}$ J). 
The collisions between the Earth and the largest meteoroids, with diameters from hundreds of meters to several kilometers, blast out the impacting objects, create wide craters, generally twenty times larger than the diameter of the meteoroids, and melt terrestrial rocks. 
Fortunately, such collisions are rare events:  statistically, 
a 100 to 200 m meteoroid hits the Earth every 1000 years, 
a 500 to 800 m meteoroid every 30000 years, 
and a 5 km meteoroid every 40 millions years.

We believe it is reasonable to assume that both the mass and linear momentum of the impacting object are negligible with respect to the Earth's mass and linear momentum, so the orbit of the Earth is not disturbed.  
Besides, the rotation period and tilt of the rotation axis of the Earth could be changed by the impact if the angular momentum of the object is large enough. 
We, however, consider impactors for which the angular momentum is at least one hundred times smaller than the Earth's; the change of the Earth's angular momentum is therefore negligible. 
The major known meteoroid impacts had such characteristics (Table \ref{table:meteoroid}).

Consequently, we reduce the extraterrestrial impact issue to the computation of the equivalent seismic magnitude corresponding to the released energy. 
The computation proposed here is based on the equations and drastic simplifications used by \citet{collins}, which are summarized below.

When an object enters the Earth's atmosphere, it loses its kinetic energy through deceleration and ablation. 
The rate of change of the velocity $v$ is given by the drag equation \citep{collins, melosh1989}:
\begin{equation}
\frac{dv}{dt}=-\frac{3\rho z C_D}{4 \rho_i L_0}v^2,
\end{equation}
where $z$ is the altitude, $C_D$ is the drag coefficient, taken equal to 2, and $\rho_i$ and $L_0$ are, 
respectively, the impactor density and diameter.
By assuming an exponential atmosphere, 
\begin{equation}
\rho(z)=\rho_0e^{-z/H}, 
\end{equation} 
where $H=$ 8 km is the scale height and $\rho_0=$1 kg/m$^3$, 
the velocity of the impactor as a function of altitude is given by:
\begin{equation}\label{vz}
v(z)=v_0 \exp\left({-\frac{3\rho(z)C_DH}{4\rho_iL_0\sin\alpha}}\right),
\end{equation}
where $\alpha$ is the entry angle
and $v_0$, the velocity at the top of the atmosphere.
On its trajectory down to the ground, the impactor goes through the increasing atmospheric pressure
and, possibly, breaks up. 
\citet{collins} have established an empirical strength-density relation
to estimate the yield strength $Y_i$ (in Pa)
\begin{equation}
\log_{10}Y_i=2.107+0.0624\sqrt{\rho_i}
\end{equation}
and give an approximate expression for the altitude of breakup $z^{\star}$:
\begin{equation}\label{z*}
z^{\star} \approx -H[\ln(\frac{Y_i}{\rho_0v_i^2})+1.308-0.314I_f-1.303\sqrt{1-I_f}],
\end{equation}
 where 
 \begin{equation} 
 I_f=4.07\frac{C_DHY_i}{\rho_iL_0v_i^2\sin\alpha},
 \end{equation} 
and $v_i$ is the impactor velocity at the surface.
 
Equation (\ref{z*}) holds provided that $I_f<1$. 
Otherwise, and more rarely, the object does not break up and the velocity at the impact is given by Equation (\ref{vz}).

Following \citet{collins}, we use the approximative pancake model \citep{chyba, melosh1981} to 
describe the disintegration of the meteoroid. 
Let us denote by $z_b$ the airburst altitude, which is the altitude of complete dispersion of the fragments. 
According to the simplifying assumptions of the pancake model, $z_b$ is given by 
\begin{equation}\label{zb}
z_b = z^{\star}-2H \ln(1+\frac{\ell}{2H}\sqrt{f_p^2-1}),
\end{equation}
where $\ell$ is the dispersion length scale:
\begin{equation}
\ell=L_0 \sin\alpha \sqrt{\frac{\rho_i}{C_D\rho(z^{\star})}}
\end{equation}
and the pancake factor $f_p$ is between 2 and 10. 
We shall adopt  \citet{collins}'s value of 7. 
If $z_b>0$, the airburst occurs in the atmosphere, there is neither impact nor associated seismic event.
If $z_b\leq0$, the fragments are not dispersed when they collide with the ground and 
the impact velocity is:
\begin{eqnarray}
v_{zr}  = v(z^{\star}) \exp & &\lbrace(-\frac{3}{4}\frac{C_D\rho(z^{\star})}{\rho_i L_0^3 \sin\alpha}\frac{H^3 L_0^2}{3 \ell^2}(32+(\frac{\ell}{H})^2 e^{z^{\star}/H} \nonumber \\
& &+ 6 e^{2 z^{\star}/H}-16 e^{3z^{\star}/2H}-3(\frac{\ell}{H})^2)\rbrace
\end{eqnarray}

The remaining kinetic energy at the moment of impact is  
\begin{equation}
E_{cr} = \frac{\pi}{12}\rho_i L_0^3 v_{zr}^2.
\end{equation}
After the impact, a fraction $k_s$ of $E_{cr}$ is radiated as seismic waves. 
Experimental data  \citep{schultz} provide $k_s \in [10^{-5}, 10^{-3}]$. 
We will take $k_s=10^{-4}$.
The seismic moment being given by 
\begin{equation}
M_0 = 2\frac{\mu}{\Delta \sigma_S}k_sE_{cr},
\end{equation}
where the stress release $\Delta \sigma_S \approx$ 3 MPa and 
the rigidity $\mu=$ 30 GPa, 
the seismic magnitude is then:
\begin{equation}
M_w = \frac{2}{3} \log_{10}(M_0)-10.73
\end{equation} 
with $M_0$ in dyn.cm (1 dyn.cm=$10^{-7}$ N/m).

We compute the magnitude $M_w$ for the different meteoroid impacts of Table \ref{table:meteoroid} 
and, in the same table, report the surface gravity perturbation associated 
with the translational excitation of the inner core. 
Note that the source is represented in terms of moment tensor by three orthogonal force couples (spherically symmetric explosion) and not as a vertical force. 
Indeed, the shock pressure would reach hundreds of gigapascals and the impact energy would vaporize the rocks and cause a spherically symmetric explosion, as observed from
the spherical shape of known craters.

\begin{figure}[!ht]
\centering
\noindent\includegraphics[width=15cm]{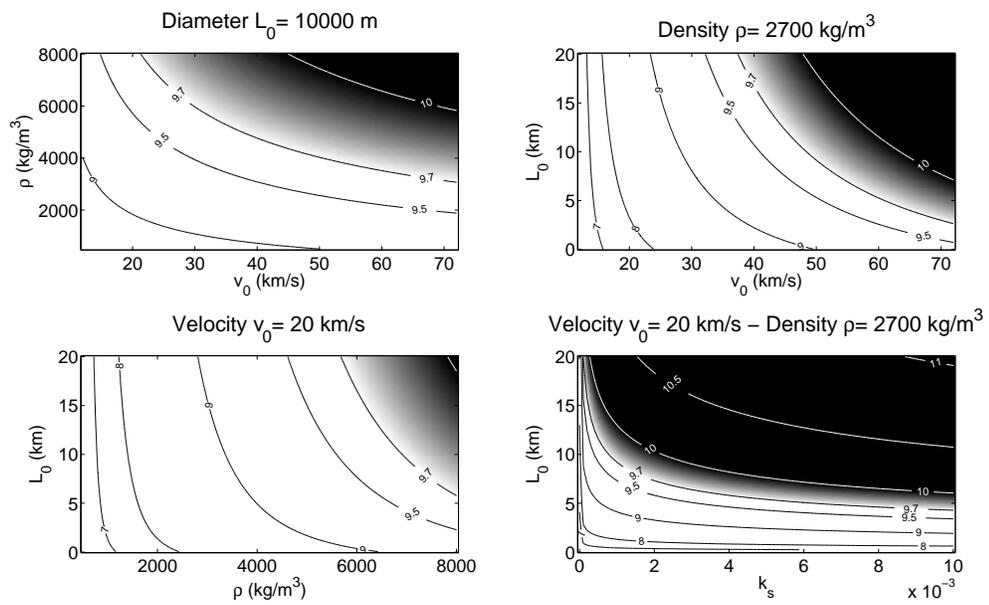}
\caption{Seismic magnitude as a function of the impactor parameters and seismic efficiency $k_s$. 
The shaded area corresponds to moment magnitudes larger than 9.7, i.e. to induced surface gravity perturbation larger than 1 nGal.}
\label{fig:impactMw}
\end{figure}

\begin{figure}[!ht]
\centering
\noindent\includegraphics[width=15cm]{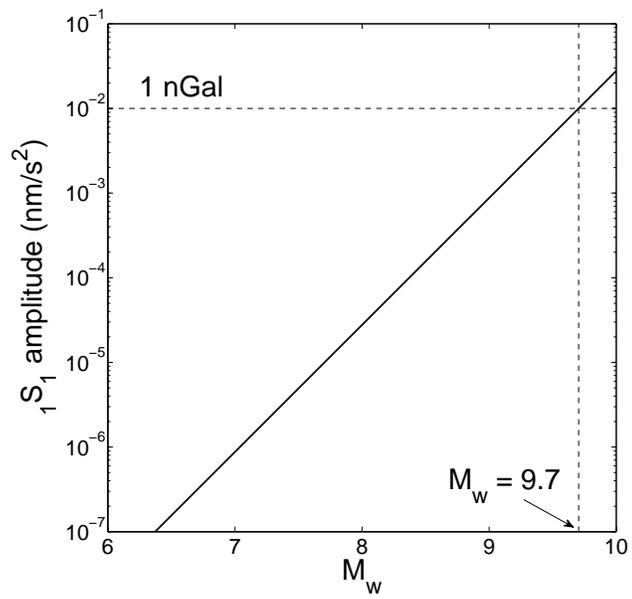}
\caption{Surface gravity perturbation induced by the Slichter mode as a 
function of the moment magnitude of a superficial energy release (explosion or object impact).}
\label{fig:1S1Ms}
\end{figure}

For the biggest meteoroid, the surface excitation amplitude of the Slichter mode
is 0.0067 nm/s$^2$, which is less than the detection threshold of 1 nGal (= 0.01 nm/s$^2$).
To determine what kind of impact would be necessary to excite sufficiently the Slichter mode so that it is 
detectable in surface gravity data, we compute the magnitude $M_w$ and gravity perturbation 
$\Delta g$ for various ranges of density $\rho_i$, velocity $v_0$, diameter $L_0$ of the object
and for seismic efficiency $k_s$ varying between $10^{-5}$ and $10^{-2}$. 
The resulting maps are plotted in Fig.\ref{fig:impactMw}. 
The shaded areas correspond to $M_w$ larger than 9.7,
which is the magnitude required for the surface gravity effect to reach the nGal detection threshold for a surficial explosive moment source (Fig. \ref{fig:1S1Ms}).

We see from Fig.\ref{fig:impactMw} that to produce a seismic event of magnitude larger than 9.7, 
the size, density or velocity of the impacting object should have unrealistic huge values. However, the value of the seismic efficiency $k_s$ 
has a direct impact on the equivalent seismic magnitude. For instance, with a seismic efficiency of $10^{-3}$ instead of $10^{-4}$, a meteoroid similar to the one 
which produced the Chicxulub crater in Mexico would be able to induce a seismic event of such a magnitude. Of course, the consequences would have been devastating.

We conclude that the surficial seismic events, including extra-terrestrial object impacts and explosions, are not efficient to make the inner-core oscillate at 
the Slichter frequency with an amplitude large enough to be observed at the surface. 
The reason is the same as for earthquakes, i.e. the excitation amplitude is directly linked to the seismic magnitude and 
the radial eigenfunctions of the Slichter mode are constant and close to zero in the mantle \citep{crossley, rogister, rosat2007}.

\section{Conclusions and perspectives}
We have investigated the excitation of the translational free motion of the inner core by a pressure due to a flow in the outer core and acting at both the ICB and CMB, 
by a surface load, which can be associated to atmospheric or oceanic loading for instance,
and by the collision between the Earth and a stellar object.
Our conclusion is that the Slichter mode would be best excited by a pressure acting at the core boundaries at time-scales shorter than half the Slichter eigenperiod. 

For the pressure source at the ICB and CMB and the loading source at the surface, 
we have considered Gaussian functions of time. 
More complicated sources should be considered, 
in particular stochastic forces produced by some turbulent flow in the core or at the surface. 
The stochastic excitation, be it oceanic and atmospheric, of normal modes has been studied for instance by \citet{tanimotoum}, 
\citet{tanimoto1999, tanimoto2007} and \citet{webb2007, webb2008}. 
However, the time scale for the Slichter mode is larger than for the other seismic normal modes, 
whose eigenperiod is shorter than 1 hour, so we should consider a theory different from the Kolmogorov theory of turbulence.

We have considered an analytic expression for the surface pressure as a source for the excitation of the Slichter mode. 
A more realistic approach should be based on actual atmospheric data 
from space correlation of worldwide barometers or from weather institutes (ECMWF, NCEP...),
provided the data are available at a time resolution higher than the Slichter period.

\newpage

%


\appendix
\section{Evaluation of the time integral I(t) (Eq. \ref{Time_Integral})}
We evaluate the following integral:
\begin{eqnarray*}
I(t)=\int_{-\infty}^{t}e^{i\nu(t-t')}e^{-[\frac{t'-T_0}{\tau}]^2} dt'
\end{eqnarray*}
We put $b=\frac{t-T_0}{\tau}$ and introduce the variable $x=\frac{t'-T_0}{\tau}$. 
$I(t)$ becomes:
\begin{eqnarray*}
I(t)=\tau \int_{-\infty}^{b}e^{i\nu \tau(b-x)}e^{-x^2}dx = \tau \int_{-\infty}^{b}e^{i\nu \tau b}e^{-x(x+i\nu\tau)}dx\\ 
\end{eqnarray*}

We perform the change of variable $y=x+i\frac{\nu\tau}{2}$:
\begin{eqnarray*}
I(t)=\tau e^{i\nu \tau b}\int_{-\infty}^{b+i\frac{\nu\tau}{2}} e^{-(y-i\frac{\nu\tau}{2})(y+i\frac{\nu\tau}{2})}dy \\
=\tau e^{i\nu \tau b}e^{-\frac{\nu^2\tau^2}{4}}\int_{-\infty}^{b+i\frac{\nu\tau}{2}} e^{-y^2}dy 
\end{eqnarray*}

The integral $\frac{2}{\sqrt \pi}\int_{z}^{+\infty}e^{-y^2}dy$ is the complementary error function ${\rm erfc}(z)=1-{\rm erf}(z)$, where 
\begin{equation}
{\rm erf}(z)=\frac{2}{\sqrt \pi}\int_{0}^{z}e^{-y^2}dy.
\end{equation}
So the time integral $I(t)$ is given by:
\begin{eqnarray*}
I(t)=\frac{\sqrt{\pi}}{2}\tau e^{i\nu(t-T_0)}e^{-\nu^2\tau^2/4}[1+ \mbox{erf}(\frac{t-T_0}{\tau}+i\frac{\nu \tau}{2})]
\end{eqnarray*}

\newpage

\bibliographystyle{elsarticle-harv}

%
%



\newpage

\begin{table}[h!]
\centering
\caption{Some meteoroid impacts on the Earth continental crust and oceanic crust. The impact angle is supposed to be 45 degrees
 and the impact velocity is 20 km/s.}
\begin{tabular}{l|c|@{} c @{}|c|@{} c @{}|@{} c @{}|}
\label{table:meteoroid}
& Date & Diameter & Density & $M_w$ & $\Delta g$ \\
Location & (AD or My BP)&  (m)&  (kg/m$^3$)&  & (nm/s$^2$)\\
\hline\hline
Tunguska Fireball & 1908 AD & 60 & 2700  & \multicolumn{2}{c|}{No impact} \\
Siberia & &  &(rock) &\multicolumn{2}{c|}{} \\
\hline
Ries Crater & $15.1 \pm 0.1$ & 1500 & 2700  & 7.4 & 3.9 $10^{-6}$ \\
Germany & &  & (rock)  &  &  \\
\hline
Rochechouart & $214 \pm 8$ & 1500 & 3350  & 7.5 &  4.9 $10^{-6}$\\
 France &  &  & (stony-iron) & &  \\
\hline
Chesapeake Bay & $35.5 \pm 0.3$ & 2300 & 2700  & 7.8 & 1.4 $10^{-5}$\\
USA & & &(rock) & &\\
\hline
Chicxulub& $64.98 \pm 0.05$ & 17500 & 2700  & 9.6 & 6.7 $10^{-3}$ \\
Mexico & & &(rock) & &
\end{tabular}
\end{table}

\end{document}